\documentclass[sigconf]{acmart}

\copyrightyear{2019}
\acmYear{2019}
\acmConference[ESEC/FSE 2019]{The 27th ACM Joint European Software Engineering Conference and Symposium on the Foundations of Software Engineering}{26--30 August, 2019}{Tallinn, Estonia}

\usepackage[keeplastbox]{flushend}

\usepackage{graphicx}

\usepackage{booktabs}
\usepackage{tabularx}
\usepackage{xspace}
\usepackage[autolanguage]{numprint}
\usepackage{hyperref}
\usepackage[inline]{enumitem}
\usepackage{multirow}
\usepackage{diagbox}
\usepackage{rotating}
\usepackage{eqparbox}
\usepackage{threeparttable}
\usepackage{array,makecell}
\usepackage{ifthen}
\usepackage[table]{xcolor}

\usepackage{tcolorbox,tikz}
\usepackage{pgfplots}
\usepackage{pgfplotstable}
\usepackage{filecontents}
\usepgfplotslibrary{statistics}
\pgfplotsset{compat=1.14} 
\usepackage{silence}
\def\cca#1{\cellcolor{black!#1}\ifnum #1>49\color{white}\fi{#1}}

\newcommand{\totalRepairToolsLiterature}{\numprint{24}\xspace}%

\newcommand{\nbBenchmarks}{5\xspace}%
\newcommand{\nbRepairTools}{11\xspace}%
\newcommand{\nbBugs}{\numprint{2141}\xspace}%
\newcommand{\nbRepairAttempts}{\numprint{23551}\xspace}%

\newcommand{\frameworkName}{\textsc{RepairThemAll}\xspace}%

\newcommand{\answer}[2]{\vspace{.1cm}{\centering\fbox{\parbox{0.97\columnwidth}{\textbf{Answer to RQ#1}. #2}}}\vspace{.2cm}}

\newboolean{showcomments}
\setboolean{showcomments}{true}

\ifthenelse{\boolean{showcomments}}
  {\newcommand{\nb}[3]{
  {\color{#2}\small\fbox{\bfseries\sffamily\scriptsize#1}}
  {\color{#2}\sffamily\small$\triangleright~$\textit{\small #3}$~\triangleleft$\GenericWarning{}{LaTeX Warning: #1: #3}}
  }
  \newcommand{\todo}[1]{{\color{red}{TODO: #1}}\GenericWarning{}{LaTeX Warning: TODO: #1}}
  }
  {\newcommand{\nb}[3]{}
  \newcommand{\todo}[1]{}
  }

\title{Empirical Review of Java Program Repair Tools}
\subtitle{A Large-Scale Experiment on \nbBugs Bugs and \nbRepairAttempts Repair Attempts}

\author{Thomas Durieux}
\affiliation{%
  \institution{INESC-ID and IST, University of Lisbon, Portugal}
}
\email{thomas@durieux.me}

\author{Fernanda Madeiral}
\affiliation{%
  \institution{Federal University of Uberl\^andia, Brazil}
}
\email{fer.madeiral@gmail.com}

\author{Matias Martinez}
\affiliation{%
  \institution{University of Valenciennes, France}
}
\email{matomartinez@gmail.com}

\author{Rui Abreu}
\affiliation{%
  \institution{INESC-ID and IST, University of Lisbon, Portugal}
}
\email{rui@computer.org}

\renewcommand{\shortauthors}{Durieux, et al.}

\begin{document}
\begin{abstract}
In the past decade, research on test-suite-based automatic program repair has grown significantly.
Each year, new approaches and implementations are featured in major software engineering venues.
However, most of those approaches are evaluated on a single benchmark of bugs, which are also rarely reproduced by other researchers.
In this paper, we present a large-scale experiment using \nbRepairTools Java test-suite-based repair tools and \nbBenchmarks benchmarks of bugs. Our goal is to have a better understanding of the current state of automatic program repair tools on a large diversity of benchmarks.
Our investigation is guided by the hypothesis that the repairability of repair tools might not be generalized across different benchmarks of bugs.
We found that the \nbRepairTools tools 1) are able to generate patches for 21\% of the bugs from the \nbBenchmarks benchmarks, and 2) have better performance on Defects4J compared to other benchmarks, by generating patches for 47\% of the bugs from Defects4J compared to 10-30\% of bugs from the other benchmarks.
Our experiment comprises \nbRepairAttempts repair attempts in total, which we used to find the causes of non-patch generation. These causes are reported in this paper, which can help repair tool designers to improve their techniques and tools.
\end{abstract}

\maketitle

\section{Introduction}

Software bugs decrease the quality of software systems from the point of view of the software system users.
Manually repairing bugs is well-known as being a difficult and time-consuming task.
To address this activity automatically, a new field of research has emerged, named \textit{automatic program repair}. Automatic program repair consists of automatically finding solutions 100\% executable to software bugs, without human intervention \cite{Monperrus2014,Monperrus2018bibliography}. The most popular approach to automatically repair bugs is to create a patch using the test suite of the program as the specification of its expected behavior. This type of approach is known  as \textit{test-suite-based program repair} \cite{Monperrus2014}, which has been applied in several repair tools in the last decade \cite{Xiong2017ACS,Yuan2018ARJA,Wen2018CapGen,Martinez2018Cardumen,White2019DeepRepair,Durieux2016DynaMoth,Saha2017ELIXIR,Le2016HDRepair,Chen2017Jaid,Martinez2016Astor,Liu2018LSRepair,Xuan2016Nopol,Durieux2017NPEFix,Kim2013PAR,Jiang2018SimFix,Hua2018SketchFix,Liu2018SOFix,Xin2017ssFix}.

The repair tools are used in \textit{empirical evaluations} so that the repairability of the repair approaches they implement is measured.
These evaluations are reported in the literature in two ways: when a new repair approach is proposed (e.g. \cite{Yuan2018ARJA}), or when a dedicated full contribution on evaluating existing repair tools is reported (e.g. \cite{Martinez2017experiment,Motwani2018evaluation,Ye2019StudyQuixBugs}).
The evaluations consist of four main aspects in general:
1) [benchmark] the selection of benchmarks of bugs;
2) [execution] the collection of data by executing repair tools on the selected bugs;
3) [observed aspect] an investigation on the effectiveness of the repair approach regarding some criteria (e.g. \textit{repairability}, \textit{correctness}, and \textit{repair time}); and finally
4) [comparison/discussion] the comparison of repair approaches and discussion.

A major problem with all previous evaluations, focusing on repair for Java programs, is that they are widely performed on the same benchmark of bugs: Defects4J \cite{Just2014Defects4J}.
In theory, this should not be a problem if Defects4J is not biased; however, no benchmark is perfect \cite{LeGoues2015ManyBugsIntroClass}.
Benchmarks should reflect the representativeness of the bugs and the projects they come from in the real world. 
The extent of the representativeness of benchmarks for real-world bugs is unknown, because even the distribution of the real world bugs is unknown. Therefore, by using a single benchmark when evaluating repair tools, a bias can be introduced, which makes hard to generalize the performance of repair tools.

In this paper, we report on a large experiment conducted on \nbRepairTools test-suite-based repair tools for Java using other benchmarks of bugs than Defects4J.
The primary goal of this experiment is to investigate if the existing repair tools behave in a similar way across different benchmarks. If a repair tool performs significantly better on one benchmark than on others, we say that the repair tool \textit{overfits the benchmark}.
The secondary goal is to understand the causes of non-patch generation from a practical view, which, to the best of our knowledge, has not been subject of investigation by the repair community.

To achieve our goals, we designed our experiment considering three out of the four main aspects usually used to evaluate repair tools:
a) on benchmark, we use \nbBenchmarks benchmarks (including Defects4J), totaling \nbBugs bugs;
b) on execution, we run \nbRepairTools repair tools on the \nbBugs bugs, using a framework we developed to automatize and simplify the execution of repair tools on different benchmarks;
c) on observed aspect, we analyze the repairability of the tools, focusing on their performance across the different benchmarks.
We do not target the fourth aspect of evaluations, which is about comparing repair tools. Our goal is not to compare repair tools among themselves, but to compare the behavior of each tool among different benchmarks.

Through our experiment, we first observed that all \nbRepairTools repair tools are able to generate a test-suite adequate patch for bugs from each of the \nbBenchmarks benchmarks.
However, when analyzing the proportion of bugs patched by the repair tools per benchmark, we found that indeed the repair tools perform better on Defects4J than on the other benchmarks.
Finally, we found six main reasons why repair tools do not succeed to generate patches for bugs. For instance, we observed that incorrect fault localization and multiple fault locations have a significant impact on patch generation. These reasons are valuable for repair tool designers and researchers to improve their tools.

To sum up, our contributions are:
\begin{itemize}
    \item A large-scale experiment of \nbRepairTools repair tools on \nbBugs bugs from \nbBenchmarks benchmarks: this is the largest study on automatic program repair ever (i.e. \nbRepairTools x \nbBugs = \nbRepairAttempts repair attempts); 
    
    \item A repair execution framework, named \frameworkName, that adds an abstraction around repair tools and benchmarks, which can be further extended to support additional repair tools and benchmarks;
    
    \item A novel study on the repairability of repair tools across multiple benchmarks, where the goal is to investigate if there exists the \textit{benchmark overfitting} problem, i.e. that the repair tools perform significantly better on the extensively used benchmark Defects4J than on other benchmarks;
    
    \item A thorough study on the non-patch generation cases on the \nbRepairAttempts repair attempts that did not result in patches.
\end{itemize}

The remainder of this paper is organized as follows.
\autoref{sec:repro:existing-evaluations-on-repair-tools} presents the literature on the evaluation of test-suite-based repair tools for Java, which grounds the motivation of our experiment.
\autoref{sec:repro:study-design} presents the design of our study, including the research questions and the data collection and analysis.
\autoref{sec:repro:results} presents the results, followed by the discussion in \autoref{sec:repro:discussion}.
Finally, \autoref{sec:repro:related-works} presents the related works, and
\autoref{sec:repro:conclusions} presents the final remarks.

\section{State of affairs on test-suite-based automatic repair tools for Java}
\label{sec:repro:existing-evaluations-on-repair-tools}

Automatic repair tools meet benchmarks of bugs when they are evaluated.
In this section, we present a review of the literature on the existing evaluations of repair tools, which is based on a two-step protocol: 1) gathering repair tools, and 2) gathering information on their evaluations, focusing on the used benchmarks and the number of bugs given as input to the repair tools.

To gather repair tools, we searched the existing living review on automatic program repair \cite{Monperrus2018living} for test-suite-based repair tools for Java, which is the focus of this work: we found \totalRepairToolsLiterature repair tools that meet this criterion.
Then, we gathered scientific papers containing evaluations of these tools.
There are two types of papers that are interesting for us: the first type consists of the presentation of a new repair approach, which also includes an evaluation conducted using a tool that implements the approach (e.g. \cite{Xuan2016Nopol}); and the second type consists of an empirical evaluation carried out on already created tools, which is a specific work  to evaluate repair tools by running them on benchmarks of bugs (e.g. \cite{Martinez2017experiment}).
We gathered 18 papers from the first type of papers (more than one tool can be presented in the same paper) and two papers from the second type.

\begin{table}[t]
    \caption{Test-suite-based program repair tools for Java.}
    \label{tab:Java-test-suite-based-repair-tools}
    \centering
    \footnotesize
    \begin{threeparttable}[b]
        \begin{tabular}{p{.145\textwidth} l r r r}
            \toprule
            \multirow{2}{*}{Repair tool} & Benchmark used & \multirow{2}{*}{\# Bugs} & \multirow{2}{*}{\# Patched\tnote{a}} & \multirow{2}{*}{\# Fixed\tnote{b}} \\
            {} & in evaluation & {} & {} & {} \\
            
            \midrule
            \multicolumn{5}{l}{\textit{\underline{Generate-and-validate}}} \vspace*{3pt} \\
            
            \rowcolor{gray!15}
            ACS \cite{Xiong2017ACS} & Defects4J & 224 & 23 & 17 \\ 
            
            \multirow{2}{*}{ARJA \cite{Yuan2018ARJA}} & Defects4J & 224 & 59 & 18 \\ 
            {} & QuixBugs \cite{Ye2019StudyQuixBugs} & 40 & 4 & 2 \\
            
            \rowcolor{gray!15}
            {} & Defects4J & 224 & 25 & 22 \\ 
            \rowcolor{gray!15}
            \multirow{-2}{*}{\textsc{CapGen} \cite{Wen2018CapGen}} & IntroClassJava & 297 & -- & 25 \\
            
            \multirow{2}{*}{Cardumen \cite{Martinez2018Cardumen}} & Defects4J & 356 & 77 & -- \\ 
            {} & QuixBugs \cite{Ye2019StudyQuixBugs} & 40 & 5 & 3 \\
            
            \rowcolor{gray!15}
            DeepRepair \cite{White2019DeepRepair} & Defects4J & 374 & 51 & -- \\
            
            \multirow{2}{*}{\textsc{Elixir} \cite{Saha2017ELIXIR}} & Defects4J & 82 & 41 & 26 \\ 
            {} & Bugs.jar & 127 & 39 & 22 \\
            
            \rowcolor{gray!15}
            GenProg-A \cite{Yuan2018ARJA} & Defects4J & 224 & 36 & -- \\
            
            HDRepair \cite{Le2016HDRepair} & Defects4J & 90 & -- & 23 \\ 
            
            \rowcolor{gray!15}
            \textsc{Jaid} \cite{Chen2017Jaid} & Defects4J & 138 & 31 & 25 \\ 
            
            \multirow{3}{*}{jGenProg \cite{Martinez2016Astor}} & Defects4J & 224 & 29 & -- \\
            {} & Defects4J \cite{Martinez2017experiment} & 224 & 27 & 5 \\
            {} & QuixBugs \cite{Ye2019StudyQuixBugs} & 40 & 2  & 0 \\
            
            \rowcolor{gray!15}
            {} & Defects4J & 224 & 22 & -- \\
            \rowcolor{gray!15}
            {} & Defects4J \cite{Martinez2017experiment} & 224 & 22 & 1 \\
            \rowcolor{gray!15}
            \multirow{-3}{*}{jKali \cite{Martinez2016Astor}} & QuixBugs \cite{Ye2019StudyQuixBugs} & 40 & 2 & 1 \\
            
            \multirow{2}{*}{jMutRepair \cite{Martinez2016Astor}} & Defects4J & 224 & 17 & -- \\
            {} & QuixBugs \cite{Ye2019StudyQuixBugs} & 40 & 3 & 1 \\
            
            \rowcolor{gray!15}
            Kali-A \cite{Yuan2018ARJA} & Defects4J & 224 & 33 & -- \\
            
            LSRepair \cite{Liu2018LSRepair} & Defects4J & 395 & 38 & 19 \\ 
            
            \rowcolor{gray!15}
            PAR \cite{Kim2013PAR} & PARDataset & 119 & 27 & -- \\ 
            
            \multirow{2}{*}{RSRepair-A \cite{Yuan2018ARJA}} & Defects4J & 224 & 44 & -- \\
            {} & QuixBugs \cite{Ye2019StudyQuixBugs} & 40 & 4 & 2 \\
            
            \rowcolor{gray!15}
            SimFix \cite{Jiang2018SimFix} & Defects4J & 357 & 56 & 34 \\ 
            
            \textsc{SketchFix} \cite{Hua2018SketchFix} & Defects4J & 357 & 26 & 19 \\ 
            
            \rowcolor{gray!15}
            SOFix \cite{Liu2018SOFix} & Defects4J & 224 & -- & 23 \\ 
            
            ssFix \cite{Xin2017ssFix} & Defects4J & 357 & 60 & 20 \\ 
            
            \rowcolor{gray!15}
            xPAR \cite{Le2016HDRepair} & Defects4J & 90 & -- & 4 \\ 
            
            \midrule
            \multicolumn{5}{l}{\textit{\underline{Semantics-driven}}} \vspace*{3pt} \\
            
            \rowcolor{gray!15}
            {} & Defects4J & 224 & 27 & -- \\
            \rowcolor{gray!15}
            \multirow{-2}{*}{DynaMoth \cite{Durieux2016DynaMoth}} & QuixBugs \cite{Ye2019StudyQuixBugs} & 40 & 2 & 1 \\
            
            \multirow{3}{*}{Nopol \cite{Xuan2016Nopol}} & ConditionDataset & 22 & 17 & 13 \\
            {} & Defects4J \cite{Martinez2017experiment} & 224 & 35 & 5 \\
            {} & QuixBugs \cite{Ye2019StudyQuixBugs} & 40 & 3 & 1 \\
            
            \midrule
            \multicolumn{5}{l}{\textit{\underline{Metaprogramming-based}}} \vspace*{3pt} \\
            
            \rowcolor{gray!15}
            {} & NPEDataset & 16 & 14 & -- \\
            \rowcolor{gray!15}
            \multirow{-2}{*}{NPEFix \cite{Durieux2017NPEFix}} & QuixBugs \cite{Ye2019StudyQuixBugs} & 40 & 2 & 1 \\
            
            \bottomrule
        \end{tabular}
        \begin{tablenotes}
            \item[a]{A \textit{patched} bug means that a repair tool fixed it with a test-suite adequate patch.}
            \item[b]{A \textit{fixed} bug means that a repair tool fixed it with a test-suite adequate patch that was confirmed to be correct.}
        \end{tablenotes}
    \end{threeparttable}
\end{table}

\autoref{tab:Java-test-suite-based-repair-tools} summarizes our review on the existing evaluations of the \totalRepairToolsLiterature repair tools based on the 20 scientific papers.
The first column presents the repair tools, which are grouped by the well-known categories \textit{generate-and-validate} and \textit{semantics-based} approaches, plus \textit{metaprogramming-based}. We named the latter category for repair tools that first create a metaprogram of the program under repair and then explore it at runtime, which in the end uses the runtime information to generate patches.

Each repair tool is associated with one or more benchmarks used in its evaluation in the table.
When a repair tool has been evaluated in more than one benchmark (or more than once in the same benchmark), we place first the benchmark used in the paper that presented the tool (i.e. first evaluation), followed by the other benchmarks with the reference for the posterior studies. For instance, in the paper that jGenProg \cite{Martinez2016Astor} is presented, there is an evaluation on Defects4J: this evaluation has no citation in the second column of the table because the evaluation is in jGenProg's paper. Later, it was evaluated again on Defects4J \cite{Martinez2017experiment} and also on QuixBugs \cite{Ye2019StudyQuixBugs}, which contain citations of the empirical evaluation papers in the table. The table also presents additional information on the evaluations, which are the number of bugs given as input to the repair tools, and the number of bugs for which the tools generated a test-suite adequate patch (i.e. patched bugs) and a correct patch (i.e. fixed bugs), reported by the gathered papers.

In total, we found 38 evaluations of the \totalRepairToolsLiterature repair tools.
Out of \totalRepairToolsLiterature, 22 repair tools were evaluated on (a subset of) bugs from Defects4J, and nine of them were recently evaluated on the QuixBugs benchmark \cite{Lin2017QuixBugs}.
In some exceptional cases, the benchmarks Bugs.jar \cite{Saha2018BugsDotjar} and IntroClassJava \cite{Durieux2016IntroClassJava} were also used.
However, the number of existing evaluations in terms of number of repair tools versus number of benchmarks is low compared to all possible combinations.
There are some benchmarks that were rarely used or never used so far: this is partially explained by the fact that some benchmarks were recently published (e.g. Bears \cite{Madeiral2019Bears}), thus they were not available when some repair tools were published.

We also observe that three repair tools were originally evaluated on datasets that were not presented in the literature in a research paper dedicated for them (i.e. PARDataset \cite{Kim2013PAR}, ConditionDataset \cite{Xuan2016Nopol} and NPEDataset \cite{Durieux2017NPEFix}). This is the case of the first evaluations of PAR, Nopol, and NPEFix. However, these repair tools were later evaluated on formally proposed benchmarks, except for PAR, which is not publicly available. PAR was later reimplemented, resulting in the tool xPAR \cite{Le2016HDRepair}, which was then evaluated on Defects4J.

\vspace{-.5em}
\section{Study Design}\label{sec:repro:study-design}

The extensive usage of Defects4J at evaluating repair tools motivates our study. Our goal is to investigate if the repair tools have a similar performance on other benchmarks of bugs.
In this section, we present the design of our study, including the research questions, the systematic selection of repair tools and benchmarks of bugs, and the data collection and analysis.

\subsection{Research Questions}

\begin{itemize}
    \item[\textbf{RQ1}.] {[Repairability]} To what extent do test-suite-based repair tools generate patches for bugs from a diversity of benchmarks?
    
    This research question guides us towards the investigation on the ability of the existing repair tools to generate test-suite adequate patches for bugs from the selected benchmarks.
    
    \item[\textbf{RQ2}.] {[Benchmark overfitting]} Is the repair tools' repairability similar across benchmarks?
    
    The repair tools have been extensively evaluated on Defects4J.
    Our goal in this research question is to investigate if they repair bugs from other benchmarks to a similar extent than they repair bugs from Defects4J.
    
    \item[\textbf{RQ3}.] {[Non-patch generation]} What are the causes that lead repair attempts to not generate patches?
    
    Existing evaluations focus on the successful cases, i.e. the bugs that a given repair tool generated patches for.
    However, to the best of our knowledge, there is no study that investigates the unsuccessful cases, i.e. a repair tool tried to repair a bug but no patch was generated.
    Our goal in this research question is to find the causes of non-patch generation so that the repair community can focus on practical limitations and improve their repair tools.
\end{itemize}

\subsection{Subject Repair Tools}

To include a Java test-suite-based program repair tool in our study, it must meet the following four inclusion criteria:

\begin{itemize}[leftmargin=*]
\item \textit{Criterion \#1.} The repair tool ought to be publicly available: our study involves the execution of repair tools, therefore tools that are not publicly available are excluded. We exclude tools with this criterion when 1) the paper where the tool was described does not include a link for the tool, 2) we cannot find the tool on the internet, and 3) we received an answer by email from the authors of the tool explaining why the tool is not available (e.g. Elixir has a confidentiality issue) or no answer at all.

\item \textit{Criterion \#2.} The repair tool ought to be possible to run: some tools are publicly available, but they are not possible to run for diverse issues (e.g. ACS uses GitHub, which recently changed its interface and does not allow programmed queries).
    
\item \textit{Criterion \#3.} The repair tool ought to be possible to run on bugs from different benchmarks beyond the one used in its original evaluation: we cannot run tools in other benchmarks if they are hardcoded to specific ones (e.g. SimFix is currently working only for Defects4J).

\item \textit{Criterion \#4.} The repair tool ought to require only the source code of the program under repair and its test suite used as oracle.
These two elements are the two inputs specified in the problem statement of test-suite-based automatic program repair \cite{Monperrus2018bibliography}.
\end{itemize}

After checking on all \totalRepairToolsLiterature repair tools presented in \autoref{tab:Java-test-suite-based-repair-tools}, we found 12 tools that meet the inclusion criteria outlined. One of them, ssFix, was further excluded because we had issues to run it, so we ended up with \nbRepairTools repair tools for our experiment.
\autoref{tab:selection-tools} presents the excluded and included tools, and for the excluded ones, it also shows the criterion they did not meet. Note that, among the included tools, we have eight generate-and-validate tools, the two semantics-based tools, and the only metaprogramming-based tool. Therefore, we cover the three approach categories. We briefly describe each selected repair tool in the remainder of this section.

\begin{table}[t]
    \caption{Selected repair tools based on our inclusion criteria.}
    \label{tab:selection-tools}
    \centering
    \small
    \begin{tabular}{@{}p{.015\textwidth} p{.23\textwidth} p{.17\textwidth}@{}}
        \toprule
        {} & Non-fulfillment Criteria & Repair Tools \\
        \midrule
        {\hspace*{-6pt}\multirow{5}{*}{\rotatebox[origin=c]{90}{\thead{Excluded\\(13)}}}} & Not public (C1) & Elixir, PAR, SOFix, xPAR \\
        {} & Not working (C2) & ACS, CapGen, DeepRepair \\
        {} & Only compatible with Defects4J (C3) & LSRepair, SimFix \\
        {} & Faulty class/method required (C4) & HDRepair, Jaid, SketchFix \\
        {} & Others & ssFix \\
        \midrule
        {\hspace*{-6pt}\rotatebox[origin=c]{90}{\thead{Included\\(\nbRepairTools)}}} & \multicolumn{2}{c}{\thead{ARJA, Cardumen, DynaMoth, jGenProg, GenProg-A, jKali, \\ Kali-A, jMutRepair, Nopol, NPEFix, RSRepair-A}} \\
        \bottomrule
    \end{tabular}
\end{table}

\textbf{jGenProg \cite{Martinez2016Astor} and GenProg-A \cite{Yuan2018ARJA}} are Java implementations of GenProg \cite{Weimer2009GenProg}, which is for C programs. GenProg is a redundancy-based repair approach \cite{Martinez2014} that generates patches using existing code (aka the \emph{ingredient}) from the system under repair, i.e., it does not synthesize new code. GenProg works at the statement level, and the repair operations are insertion, removal, and replacement of statements.

\textbf{jKali \cite{Martinez2016Astor} and Kali-A \cite{Yuan2018ARJA}} are Java implementations of Kali \cite{Qi2015Kali}.
Kali was conceived to show that most of the patches synthesized by GenProg over the ManyBugs benchmark \cite{LeGoues2012study} consist of avoiding the execution of code. 
The operators implemented in Kali are removal of statements, modification of \texttt{if} conditions to \textit{true} and \textit{false}, and insertion of \texttt{return} statements.

\textbf{jMutRepair \cite{Martinez2016Astor}} is an implementation of the mutation-based repair approach presented by Debroy and Wong~\cite{Debroy2010MutRepair} for Java. It considers three kinds of mutation operators: relational (e.g. $==$), logical (e.g. $\&\&$) and unary (i.e. addition or removal of the negation operator $!$). jMutRepair performs mutations on those operators in suspicious \texttt{if} condition statements.

\textbf{Nopol \cite{Xuan2016Nopol}} is a semantics-based repair tool dedicated to repair buggy \texttt{if} conditions and to add missing \texttt{if} preconditions.
Nopol uses the so-called angelic values to determine the expected behavior of suspicious statements: an angelic value is an arbitrary value that makes all failing test cases from the program under repair pass.
Nopol then collects those values at runtime and encodes them into a Satisfiability Modulo Theory (SMT) formula to find an expression that matches the behavior of the angelic value.
When the SMT formula is satisfiable, Nopol translates the SMT solution into a source code patch.

\textbf{DynaMoth \cite{Durieux2016DynaMoth}} is a repair tool integrated into Nopol that also targets buggy and missing \texttt{if} conditions.
The difference between DynaMoth and Nopol is that instead of using an SMT formula to generate the patch, it uses the Java Debug Interface to access the runtime context and collects variable and method calls.
Then, DynaMoth combines those variables and method calls to generate more complex expressions until it finds one that has the expected behavior.
This allows the generation of patches containing method calls with parameters, for instance.

\textbf{NPEFix \cite{Durieux2017NPEFix}} is different from the generate-and-validate and semantics-based tools, it is a metaprogramming-based tool. 
It means that NPEFix modifies the program under repair to include several repair strategies that can be activated during the runtime.
NPEFix repairs programs that crash due to a null pointer exception. 
NPEFix runs the failing test-case several times and activates a different repair strategy for each execution. In the end, knowing the repair strategies that have worked, together with information of the context that they worked, a patch is created. 
Note that NPEFix works in a similar way than semantics-based tools in this last step: if a patch is found, it means the patch is already satisfactory.

\textbf{ARJA \cite{Yuan2018ARJA}} is a genetic programming approach that optimizes the exploration of the search space by combining three different approaches: a patch representation for decoupling properly the search subspaces of likely-buggy locations, operation types and ingredient statements;
a multi-objective search optimization for minimizing the weighted failure rate and for searching simpler patches; 
and a method/variable scope matching for filtering the replacement/inserted code to improve compilation rate.

\textbf{Cardumen \cite{Martinez2018Cardumen}} is a test-suite-based repair approach that works at the level of expressions.
It synthesizes new expressions (that are used to replace suspicious expressions) as follows.
First, it mines templates (i.e., piece of code at the level of expression, where the variables are replaced by placeholders) from the code under repair.
Then, for creating a candidate patch that replaces a suspicious expression $se$, Cardumen selects a compatible template (i.e. the evaluation of the template and the $se$ return compatible types) and creates a new expression from it by replacing all its placeholders with variables frequently used in the context of $se$.

\textbf{RSRepair-A \cite{Yuan2018ARJA}} is a Java implementation of RSRepair \cite{Qi2014RSRepair}.
RSRepair repairs is a test-suite-based repair approach for C that has been created to compare the performance between genetic
programming (GenProg) and random search in the case of automatic program repair.
It showed that in 23/24 RSRepair finds patches faster than GenProg.

\subsection{Subject Benchmarks of Bugs}

To select benchmarks of bugs for our study, we defined the following three inclusion criteria:

\begin{itemize}[leftmargin=*]
\item \textit{Criterion \#1}: The benchmark must contain bugs in the Java language: this criterion excludes benchmarks such as ManyBugs \cite{LeGoues2015ManyBugsIntroClass}, IntroClass \cite{LeGoues2015ManyBugsIntroClass}, Codeflaws \cite{Tan2017Codeflaws} and BugsJS \cite{Gyimesi2019BugsJS}.

\item \textit{Criterion \#2}: The benchmark must be peer-reviewed, presented in the literature in a research paper dedicated for it: this criterion excludes benchmarks such as PARDataset \cite{Kim2013PAR}, ConditionDataset \cite{Xuan2016Nopol}, and NPEDataset \cite{Durieux2017NPEFix}.

\item \textit{Criterion \#3}: The benchmark must include, for each bug, at least one failing test case: this criterion excludes benchmarks such as iBugs \cite{Dallmeier2007iBugs}.
\end{itemize}

After searching the literature for benchmarks that meet our criteria, we ended up with \nbBenchmarks benchmarks. \autoref{tab:repro:benchsize} summarizes them by presenting their sizes in number of projects, bugs and lines of code. We present a brief description of them as follows.

\begin{table}[t]
    \caption{Selected benchmarks of bugs and their sizes.}
    \label{tab:repro:benchsize}
    \centering
    \small
    \begin{tabular}{@{}l r r r@{}}
        \toprule
        Benchmark                                    & \# Projects & \# Bugs & LOC (Java) \\
        \midrule
        Bears \cite{Madeiral2019Bears}                &         72 &    251 & \numprint{62597}\\
        Bugs.jar \cite{Saha2018BugsDotjar}                      &          8 &   1158 & \numprint{212889}\\
        Defects4J \cite{Just2014Defects4J}                   &          6 &    395 & \numprint{129592}\\
        IntroClassJava \cite{Durieux2016IntroClassJava} &        6 &    297 & \numprint{230}\\
        QuixBugs \cite{Lin2017QuixBugs}                       &         40 &     40 & \numprint{190}\\
        \midrule
        Total                                         &        130 &   \nbBugs & \numprint{146428}\\
        \bottomrule
    \end{tabular}
\end{table}

\textbf{Defects4J} \cite{Just2014Defects4J} contains 395 bugs from six widely used open source Java projects with an average size of \numprint{129592} lines of Java code.
The bugs were extracted by the identification of bug fixing commits with the support of the bug tracking system and the execution of tests on the bug fixing program version and its reverse patch (buggy version).
Despite the fact this benchmark was first proposed to the software testing community, it has been used for several works on automatic program repair.

\textbf{Bugs.jar} \cite{Saha2018BugsDotjar} contains \numprint{1158} bugs from eight Apache projects with an average size of \numprint{212889} lines of Java code. 
It was created using the same strategy than Defects4J. Its main contribution is the high number of bugs.

\textbf{Bears} \cite{Madeiral2019Bears} contains 251 bugs from 72 different GitHub projects with an average size of \numprint{62597} lines of Java code.  
It was created by mining software repositories based on commit building state from Travis Continuous Integration.
Bears has the largest diversity of project compared to previous bug benchmarks.

\textbf{IntroClassJava} \cite{Durieux2016IntroClassJava} contains 297 bugs from six different student projects.
It is a transpiled version to Java of the bugs from the C benchmark IntroClass \cite{LeGoues2015ManyBugsIntroClass}.
In the transpiled version, the projects have on average 230 lines of code.

\textbf{QuixBugs} \cite{Lin2017QuixBugs} contains 40  single line bugs from 40 programs, which are translated into both Java and Python languages.
Each program corresponds to the implementation of one algorithm such as Quicksort and contains on average 190 lines of code.
This is the first multi-lingual program repair benchmark.

\subsection{Building a Repair Execution Framework}\label{sec:repro:runningrepair}

\begin{figure}[t]
    \centering
    \includegraphics[width=0.94\columnwidth]{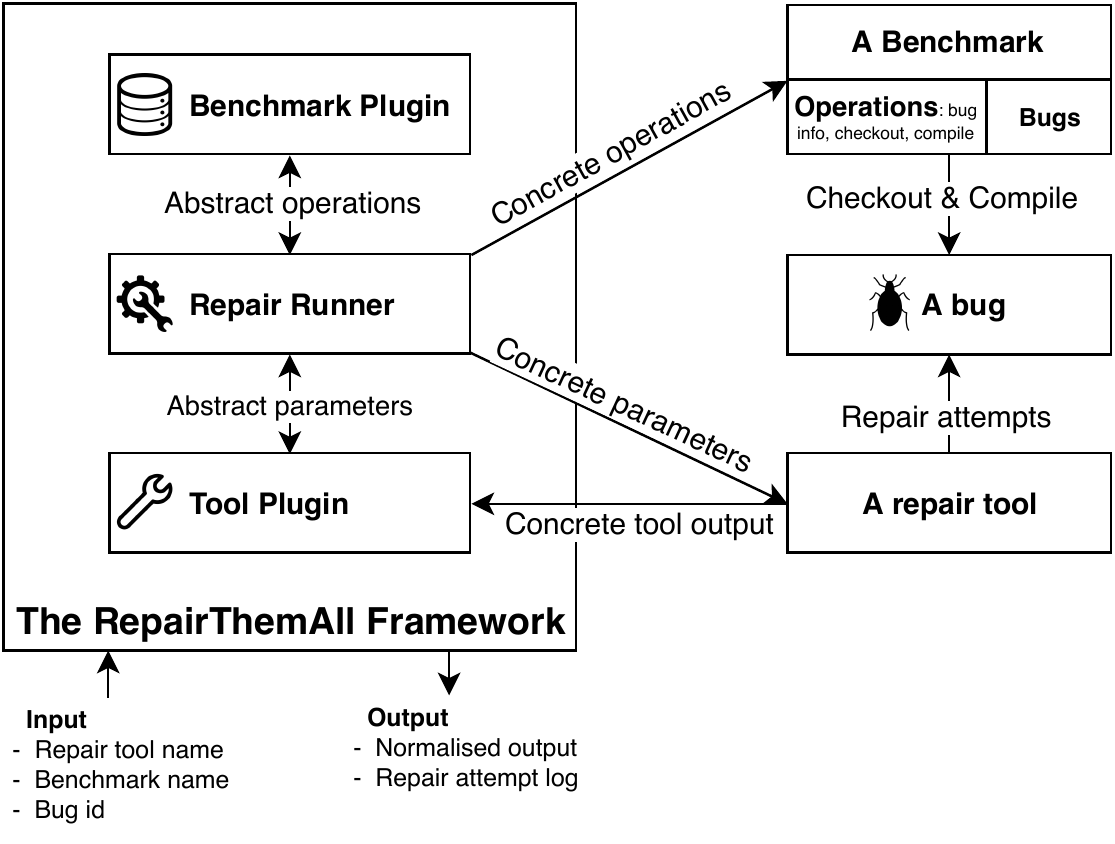}
    \caption{The \frameworkName framework.}
    \label{fig:repro:RepairThemAll-framework}
\end{figure}

To run the repair tools on different benchmarks of bugs, we created an execution framework named \frameworkName, which provides an abstraction around repair tools and benchmarks.
\autoref{fig:repro:RepairThemAll-framework} illustrates the overview of the framework.
It is composed of three main components:
1) \textit{repair tool plug-in}, where there is the abstraction around repair tools, allowing the addition and removal of tools,
2) \textit{benchmark plug-in}, where there is the abstraction around benchmarks, also allowing the addition and removal of benchmarks,
and 3) \textit{repair runner}, which works as a fa\c{c}ade for the execution of repair tools on specific bugs.

For the creation of the framework, we performed three main tasks, one for each component.
First, for \textit{repair tool plug-in}, we identify the common parameters that are required by the repair tools, which we refer to as \textit{abstract parameters}. We then map these abstract parameters to the \textit{actual parameters} of each repair tool. This is necessary because the repair tools use different parameter names and input formats. For instance, an abstract parameter is the source code folder path, and the actual parameter for source code folder path in ARJA is \texttt{DsrcJavaDir}, but in jGenProg is \texttt{srcjavafolder}.
We identify eight common parameters for the repair tools:
\begin{enumerate*}
    \item source code path,
    \item test path,
    \item binary path of the source code,
    \item binary path of the tests,
    \item the classpath,
    \item the java version (compliance level),
    \item the failing test class name, and
    \item workspace directory.
\end{enumerate*}
\frameworkName also supports the setting of actual parameters existing in the repair tools to tune specific executions, i.e., actual parameters that are not mapped to any abstract parameter.

Then, for \textit{benchmark plug-in}, we identify the \textit{abstract operations} that should be performed to use the bugs from benchmarks (e.g. check out a buggy program given a bug id).
We map these abstract operations to the \textit{actual operations} of each benchmark when they are available.
We define three required bug usage operations to be able to use the benchmark with repair tools:
\begin{enumerate*}
    \item to check out a specific bug (buggy source code files) at a given location,
    \item to compile the buggy source code and the tests, and
    \item to provide information on the bug to be given as input to repair tools (i.e. the eight parameters previously mentioned, except workspace directory). If the bug is from a multi-module project, the source code and test paths should be related to the module that contains the bug.
\end{enumerate*}
Only Defects4J provides bug usage operations that fully covers the required abstract operations. Consequently, we had to build above the four other benchmarks the missing operations, e.g. to check out a bug from the QuixBugs benchmark.

Finally, for the \textit{repair runner}, we design the input and output in a simple way so that one can easily interact with the \frameworkName framework as well as interpret the results.
For the input, one can start an execution of repair tools on benchmarks as a simple command line: for instance, the command \texttt{./repair.py Nopol --benchmark Defects4J --id Chart-7} starts the execution of Nopol on the bug Chart 7 of Defects4J.
At the end of this execution, the repair runner generates a standardized output divided into two files: the log of the repair attempt execution (\texttt{repair.log}),
and the normalized JSON file (\texttt{results.json}) containing the location of the patches generated by the tools, if any, and the textual difference between the patches and their buggy program versions.
This standardized output is due to the abstraction around the output format (output normalization) we create to simplify the analysis and the readability of the results from the different repair tools.

The \frameworkName framework currently contains \nbRepairTools repair tools and \nbBenchmarks benchmarks of bugs, and it allows the plug-in of other ones to help the repair community to compare different approaches.
Moreover, the framework allows users to set repair tool executions such as the timeout and the limit of generated patches. It is publicly available at \cite{RepairThemAll}, which includes a tutorial with all the steps to use it and to integrate new repair tools and new benchmarks.

\vspace{-1em}
\subsection{Data Collection and Analysis}

To answer our research questions, we executed the \nbRepairTools repair tools on the \nbBenchmarks benchmarks using \frameworkName, resulting in patches that are further used for analysis.
In this section, we describe the repair tools' setup (\autoref{sec:repro:setup}) and their execution (\autoref{sec:repro:execution}), and the analysis we performed on the repair attempts to determine the possible causes of non-patch generation (\autoref{sec:repro:patchanalysis}).

\subsubsection{Repair tools' setup}\label{sec:repro:setup}

For this experiment, we set the time budget to two hours per repair attempt: a repair attempt consists of the execution of one repair tool over one bug.
We also configured the repair tools to stop the execution of repair attempts when they already generated one patch.
However, ARJA, GenProg-A, Kali-A, RSRepair-A, and NPEFix do not have this option, they stop their repair attempts when they consume their own tentative budget, or by timeout.
Moreover, we configured repair tools to run on one predefined random seed: due to the huge computational power required for this experiment, we were not able to run the repair tools with additional seeds.
Finally, \autoref{tab:repro:toolversion} presents the version of each repair tool that we used in this study.
The logs of the repair attempts are available at \cite{Logs}.

\begin{table}
    \caption{The used version of each repair tool.}
    \label{tab:repro:toolversion}
    \centering
    \small
    \begin{tabularx}{1\columnwidth}{@{}X l l l@{}}
        \toprule
        Repair tool & Framework & GitHub repository & Commit id \\
        \midrule
        ARJA, GenProg-A, Kali-A, RSRepair-A  & \multirow{2}{*}{\textsc{Arja}} & \multirow{2}{*}{\href{https://github.com/yyxhdy/arja}{yyxhdy/arja}}          & \multirow{2}{*}{e60b990f9} \\
        \midrule
        {Cardumen, jGenProg, jKali, jMutRepair} & \multirow{2}{*}{\textsc{Astor}} & \multirow{2}{*}{\href{https://github.com/SpoonLabs/astor/}{SpoonLabs/astor}} & \multirow{2}{*}{26ee3dfc8} \\
        \midrule
        DynaMoth, Nopol                       & \textsc{Nopol} & \href{https://github.com/SpoonLabs/nopol}{SpoonLabs/nopol}                   & 7ba58a78d \\
        \midrule
        NPEFix                                & -- & \href{https://github.com/Spirals-Team/npefix}{Spirals-Team/npefix}           & 403445b9a \\
        \bottomrule
    \end{tabularx}
\end{table}

\subsubsection{Large scale execution}\label{sec:repro:execution}

To our knowledge, our experimental setup is the biggest one on patch generation studies, in terms of number of repair tools and bugs, and also in execution time.
In total, we executed \nbRepairTools repair tools on \nbBugs bugs from 130 open-source projects that the selected \nbBenchmarks benchmarks provide.
This represents \nbRepairAttempts repair attempts, which took 314 days and 12.5 hours of combined execution, almost a year of continuous execution.
This experiment would not be possible without the support of the cluster Grid'5000 \cite{grid5000} that provided us the required computing power to conduct this work.

\subsubsection{Finding causes of non-patch generation}\label{sec:repro:patchanalysis}

Prior studies on patch generation mainly focused on the ability of approaches to generate patches, and do not investigate the reasons why non-patch generation happens.
The study on non-patch generation is important to make research progress so that authors of repair tools can improve their tools.
Since there is a lack of knowledge on that subject, we are not able to automatically perform the detection of the reasons why patches are not generated for bugs, and therefore manual analysis is required.
Due to the scale of our experiment setup including \nbRepairAttempts different patch generation attempts, it is unrealistic to manually analyze each attempt log to understand what happened. 
We identify the major causes of non-patch generation by analyzing a sample of the repair attempt logs.
We do not predefine the sample because we observe during preliminary investigation that identical behaviors happen for groups of repair attempts. For instance, we found that for all bugs from a specific project, all the repair tools have the same issue in the fault localization. For that reason, predefining a sample is not optimal, because we would analyze attempt logs that we already know what is the cause of non-patch generation.

\vspace{-.5em}
\section{Results}\label{sec:repro:results}

The results of our empirical study, as well as the answers to our research questions, are presented in this section.

\subsection{[RQ1] Repairability of the \nbRepairTools Repair Tools}

In this research question, we analyze the repairability of the \nbRepairTools repair tools on the total of \nbBugs bugs. For that, we calculated the number of patched bugs and also the number of bugs that are commonly patched by tool.

\autoref{fig:repro:repairability} presents the repairability of the repair tools in descending order by the number of patched bugs.
For each tool, it shows the number of unique bugs patched by the tool (dark grey), the number of patched bugs that other repair tools also patched (light grey), and the total number of patched bugs with the proportion over all \nbBugs included in this study.
For instance, Nopol synthesizes patches for 213 bugs in total (9.9\% of all bugs), where 57 are uniquely patched by Nopol, and 156 are patched by Nopol and other tools.

We observe that Nopol, DynaMoth, and ARJA are the three tools that generate test-suite adequate patches for the highest number of bugs, with respectively 213, 206 and 146 patched bugs in total.
NPEFix, on the other hand, generates patches for the fewest number of bugs (15). It can be explained by the narrow repair scope of this tool, i.e. bugs exposed by null pointer exception.

On bugs uniquely patched by tools, we observe that only jKali failed to generate patches for bugs that are not patched by other tools, and that DynaMoth is the tool that patches the highest number of unique bugs.
However, NPEFix is the tool that has the highest proportion of unique patched bugs, i.e. 53\% of the 15 bugs patched by NPEFix are unique.

\pgfplotstableread[col sep=comma, header=true]{repairability.csv}{\datatable}
\pgfplotstablegetrowsof{\datatable}
\edef\numberofrows{\pgfplotsretval}

\begin{figure}[t]
  \centering
  \small
  \begin{tikzpicture}
  \pgfplotsset{
        show sum on top/.style={
            /pgfplots/scatter/@post marker code/.append code={%
                \node[
                    at={(normalized axis cs:%
                            \pgfkeysvalueof{/data point/x},%
                            \pgfkeysvalueof{/data point/y})%
                    },
                    anchor=west,
                ]
                {\pgfkeys{/pgf/fpu=true}\pgfmathprintnumber{\pgfkeysvalueof{/data point/x}}~(\pgfmathparse{\pgfkeysvalueof{/data point/x}/2141*100}%
    \pgfmathprintnumber[fixed,precision=1]{\pgfmathresult}\%)};
            },
        },
    }
    \begin{axis}
    [xbar stacked,
    width=1\linewidth,
    height=5.5cm,
    bar width=6pt,
	xlabel=\# Patched bugs,
    xlabel near ticks,
    xmin=0,
    xmax=265,
    ytick=data,
    yticklabels from table={\datatable}{tool},
    yticklabel style={align=right, text width=1.32cm},
    y dir=reverse,
    enlarge y limits=0.06,
    xtick pos=left,
    ytick pos=left,
    legend pos=south east,
    legend cell align={left},
    legend image code/.code={
        \draw [#1] (0cm,-0.1cm) rectangle (0.3cm,0.07cm);
    }
    ]
    
    \addplot[bar shift=0pt, draw=black, fill=black!25, point meta=explicit, nodes near coords] table[x=unique, y expr=\coordindex, meta index=1]{\datatable};
    
    \addplot[bar shift=0pt, draw=black, fill=black!10, point meta=explicit, nodes near coords,show sum on top] table[x=overlapped, y expr=\coordindex, meta index=2]{\datatable};
    
    \legend{Unique,Overlapped}
    \end{axis}
  \end{tikzpicture}
  \caption[Repairability of the \nbRepairTools repair tools on \nbBugs bugs.]{Repairability of the \nbRepairTools repair tools on \nbBugs bugs.\footnotemark}
  \label{fig:repro:repairability}
\end{figure}
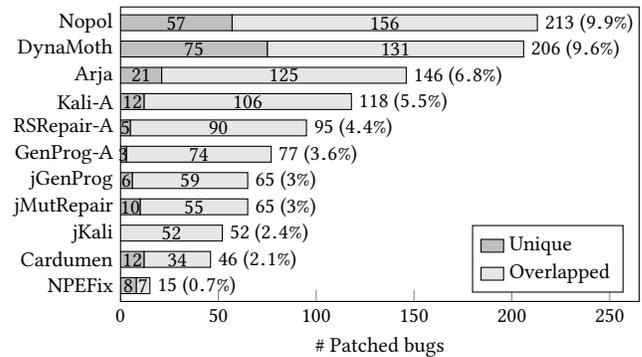
\footnotetext{The full list of the patched bugs and the textual patches are available in \cite{Website}.}

\begin{table*}[t]
    \caption{The number of overlapped patched bugs per repair tool.
    Each row $r$ presents the percentage of overlapped patched bugs of one tool  $t_r$ with the rest of the tools. For instance,  45\% of the bugs patched by ARJA (row 2) are also patched by GenProg-A (column 3). On the contrary,  85\% of the bugs patched by GenProg-A (row 3) are also patched by ARJA (column 2).}
    \label{tab:repro:commonpatch}
    \centering
    \small
    \begin{tabular}{@{}l|r r r r r r r r r r r}
    \toprule
            & ARJA        & GenProg-A   & Kali-A      & RSRepair-A  & Cardumen    & jGenProg    & jKali       & jMutRepair  & Nopol       & DynaMoth    & NPEFix       \\\midrule
 ARJA       & \textbf{13\% (20)} & \cca{45}\% (66) & \cca{56}\% (82) & \cca{55}\% (81) & \cca{15}\% (23) & \cca{30}\% (44) & \cca{27}\% (40) & \cca{19}\% (29) & \cca{36}\% \enspace(53) & \cca{32}\% \enspace(48) & \cca{2}\% (4)  \\
 GenProg-A  & \cca{85}\% (66) & \textbf{3\% \enspace(3)} & \cca{63}\% (49) & \cca{81}\% (63) & \cca{22}\% (17) & \cca{40}\% (31) & \cca{37}\% (29) & \cca{23}\% (18) & \cca{42}\% \enspace(33) & \cca{40}\% \enspace(31) & \cca{2}\% (2)  \\
 Kali-A     & \cca{69}\% (82) & \cca{41}\% (49) & \textbf{9\% (11)} & \cca{46}\% (55) & \cca{16}\% (20) & \cca{28}\% (34) & \cca{37}\% (44) & \cca{23}\% (28) & \cca{47}\% \enspace(56) & \cca{45}\% \enspace(54) & \cca{1}\% (2)  \\
 RSRepair-A & \cca{85}\% (81) & \cca{66}\% (63) & \cca{57}\% (55) & \textbf{5\% \enspace(5)} & \cca{17}\% (17) & \cca{38}\% (37) & \cca{31}\% (30) & \cca{21}\% (20) & \cca{37}\% \enspace(36) & \cca{36}\% \enspace(35) & \cca{2}\% (2)  \\
 Cardumen   & \cca{50}\% (23) & \cca{36}\% (17) & \cca{43}\% (20) & \cca{36}\% (17) & \textbf{26\% (12)} & \cca{65}\% (30) & \cca{45}\% (21) & \cca{32}\% (15) & \cca{21}\% \enspace(10) & \cca{26}\% \enspace(12) & \cca{4}\% (2)  \\
 jGenProg   & \cca{67}\% (44) & \cca{47}\% (31) & \cca{52}\% (34) & \cca{56}\% (37) & \cca{46}\% (30) & \textbf{9\% \enspace(6)} & \cca{55}\% (36) & \cca{41}\% (27) & \cca{29}\% \enspace(19) & \cca{36}\% \enspace(24) & \cca{3}\% (2)  \\
 jKali      & \cca{76}\% (40) & \cca{55}\% (29) & \cca{84}\% (44) & \cca{57}\% (30) & \cca{40}\% (21) & \cca{69}\% (36) & \textbf{0\% \enspace(0)} & \cca{57}\% (30) & \cca{53}\% \enspace(28) & \cca{67}\% \enspace(35) & \cca{1}\% (1)  \\
 jMutRepair & \cca{44}\% (29) & \cca{27}\% (18) & \cca{43}\% (28) & \cca{30}\% (20) & \cca{23}\% (15) & \cca{41}\% (27) & \cca{46}\% (30) & \textbf{15\% (10)} & \cca{58}\% \enspace(38) & \cca{30}\% \enspace(20) & \cca{1}\% (1)  \\
 Nopol      & \cca{24}\% (53) & \cca{15}\% (33) & \cca{26}\% (56) & \cca{16}\% (36) & \cca{4}\% (10) & \cca{8}\% (19) & \cca{13}\% (28) & \cca{17}\% (38) & \textbf{26\% \enspace(57)} & \cca{53}\% (114) & $<$\cca{1}\% (2)  \\
 DynaMoth   & \cca{23}\% (48) & \cca{15}\% (31) & \cca{26}\% (54) & \cca{16}\% (35) & \cca{5}\% (12) & \cca{11}\% (24) & \cca{16}\% (35) & \cca{9}\% (20) & \cca{55}\% (114) & \textbf{36\% \enspace(75)} & $<$\cca{1}\% (1)  \\
 NPEFix     & \cca{26}\% \enspace(4) & \cca{13}\% \enspace(2) & \cca{13}\% \enspace(2) & \cca{13}\% \enspace(2) & \cca{13}\% \enspace(2) & \cca{13}\% \enspace\enspace(2) & \cca{6}\% \enspace(1) & \cca{6}\% \enspace(1) & \cca{13}\% \enspace(2) & \cca{6}\% \enspace\enspace(1) & \textbf{53\% (8)}  \\
 \bottomrule
\end{tabular}
\end{table*}

The overlapping between each pair of repair tools in number of bugs is presented in \autoref{tab:repro:commonpatch}. 
In the case where the column name and the line name are the same (main diagonal), it presents the number of uniquely bugs patched by the tool.
For instance, 20 bugs have been uniquely patched by ARJA, which repairs other 66 bugs that are also patched by GenProg-A.

We observe a large overlap between repair tools that share the same patch generation framework, i.e. the framework where the repair tools are implemented (see \autoref{tab:repro:toolversion}).
For instance, ARJA has an overlap of 45\% with GenProg-A, 56\% with Kali-A, and 55\% with RSRepair-A, all implemented in the \textsc{Arja} framework.
However, ARJA has an overlap ranging from 2\% to 36\% with the other repair tools.
DynaMoth has an overlap of 55\% with Nopol, but only 0\% to 26\% with the other tools.
Each tool implemented in the \textsc{Astor} framework (e.g. jGenProg) has a big overlap with other tools in \textsc{Astor}. Moreover, the tools in \textsc{Astor} also present high overlapping with the tools in the \textsc{Arja} framework, which are tools sharing similar repair approaches.

\answer{1}{
\textbf{To what extent do test-suite-based repair tools generate patches for bugs from a diversity of benchmarks?}
The \nbRepairTools repair tools are able to generate patches for bugs ranging from 15 to 213 bugs, from a total of \nbBugs bugs.
They are complementary to each other, because 10/\nbRepairTools repair tools fix unique bugs (all but jKali).
We also observe that the overlapped repairability of the tools is impacted by their similar implemented repair approaches, and also by the patch generation framework where they are implemented.
}

\subsection{[RQ2] Benchmark Overfitting}

In this research question, we compare the repairability of the repair tools on the bugs from the extensively used benchmark Defects4J with their repairability on the other benchmarks included in this study, which are Bears, Bugs.jar, IntroClassJava, and QuixBugs.


\autoref{tab:repro:nbgeneratedpatches} shows the number of bugs that have been patched by each repair tool per benchmark.
We first observe that Defects4J is the benchmark with the highest number of unique patched bugs (187), which represents 47.34\% of all Defects4J bugs. 
The next most patched benchmarks are QuixBugs with 30\% and IntroClassJava with 20.87\% of their bugs.
This difference can also be observed in the total number of generated patches per benchmark: Defects4J is still dominating the ranking with 550 generated patches, even that it contains fewer bugs than Bugs.jar (395 versus \numprint{1158} bugs).

To test if the repairability of the repair tools
is independent of Defects4J, we applied the Chi-square test on the number of patched bugs for Defects4J compared to the other benchmarks.
The null hypothesis of our test is that \textit{the number of patched bugs by a given tool is independent of Defects4J}.
We observed in \autoref{tab:repro:nbgeneratedpatches} that the \textit{p-value} is smaller than the significance level \textit{$\alpha$ < .05} for all repair tools.
Hence, we reject the null hypothesis for those \nbRepairTools tools, and we conclude that the number of patched bugs by them is dependent of Defects4J.
Therefore, \textit{repair tools overfit Defects4J}.

The repairability of the repair tools on Defects4J cannot be only explained by the repair approaches.
We raised three hypotheses that can potentially explain the repairability difference between Defects4J and the other benchmarks: 1) there is a technical problem in the repair tools, 2) the bug fix isolation performed on Defects4J has an impact on repairing Defects4J bugs, and 3) the distribution of the bug types in Defects4J is different from the other benchmarks.

\textbf{1. [Technical problems in the repair tools]}
In RQ1, we observed the importance of the implementation of the tools for the repairability.
One hypothesis that can explain the fact that repair tools overfit Defects4J is that the authors of the repair tools have debugged and tuned their frameworks for Defects4J and, consequently, improved significantly the repairability of their tools for this specific benchmark.
For instance, they may have paid attention to not let the dependencies of the repair tools to interfere with the classpath of the Defects4J bugs, in order to preserve the behavior of test executions on the Defects4J bugs.
However, this issue can affect the bugs of other benchmarks.

\textbf{2. [Bug fix isolation performed on Defects4J]}
The second hypothesis is related to the way that Defects4J has been created. A bug fixing commit might include other changes that are not related to the actual bug fix. Then given a bug fixing commit, the authors of Defects4J recreated the buggy and patched program versions so that the diff between the two versions contains only changes related to the bug fix: this is called bug fix isolation.
The resulted isolated bug fixes facilitate studies on patches~\cite{Sobreira2018defects4jdissection}.
However, such a procedure can potentially have an impact on the repairability of the repair tools.
For instance, by comparing the developer patch \cite{HumanIsolatedPatch} with the Defects4J patch \cite{Defects4JIsolatedPatch} for the bug Closure-51, we observe that the method \texttt{isNegativeZero} has been introduced in the buggy program version, which contains part of the logic for fixing the bug. 
The presence of this method in the buggy program version can simplify the generation of patches by the repair tools or introduce an ingredient for genetic programming repair approaches.

\textbf{3. [Bug type distribution in the benchmarks]}
Our final hypothesis is related to the distribution of the bugs in the different benchmarks.
Defects4J might contain more bugs that can be patched by the repair tools compared to the other benchmarks.
For that reason, the bug type distribution of each benchmark should be further analyzed and correlated with the repairability of the tools.
\begin{table*}[t]
    \caption{Number of bugs patched by at least one test-adequate patch and the p-value of the Chi-square test of independence between the number of patched bugs from Defects4J compared to the other benchmarks.}
    \label{tab:repro:nbgeneratedpatches}
    \centering
    \small
    \begin{tabular}{l|r r r r r|r|r@{}}
    \toprule
    \diagbox[height=5.2em,width=7em]{Repair tool}{Benchmark} & \thead{Bears (251)} & \thead{Bugs.jar (\numprint{1158})} & \thead{Defects4J (395)} & \thead{IntroClassJava (297)} & 
    \thead{QuixBugs (40)} & \thead{Total (\nbBugs)} & p-value \\
    \midrule
 ARJA         & 12 \enspace(4\%) & 21 \enspace(1\%) & 86       (21\%) & 23 \enspace(7\%) & 4        (10\%) & 146 \enspace(7\%)& $< 0.00001$ \\
 GenProg-A    & 1         (<1\%) & 9        (<1\%)  & 45       (11\%) & 18 \enspace(6\%) & 4        (10\%) & 77 \enspace(3\%) & $< 0.00001$ \\
 Kali-A       & 15 \enspace(5\%) & 24 \enspace(2\%) & 72       (18\%) & 5  \enspace(1\%) & 2 \enspace(5\%) & 118 \enspace(5\%)& $< 0.00001$ \\
 RSRepair-A   & 1         (<1\%) & 6        (<1\%)  & 62       (15\%) & 22 \enspace(7\%) & 4        (10\%) & 95 \enspace(4\%) & $< 0.00001$ \\
 Cardumen     & 13 \enspace(5\%) & 12 \enspace(1\%) & 17 \enspace(4\%)& 0  \enspace(0\%) & 4        (10\%) & 46 \enspace(2\%) &   $0.00107$ \\
 jGenProg     & 13 \enspace(5\%) & 14 \enspace(1\%) & 31 \enspace(7\%)& 4  \enspace(1\%) & 3 \enspace(7\%) & 65 \enspace(3\%) & $< 0.00001$ \\
 jKali        & 10 \enspace(3\%) & 8        (<1\%)  & 27 \enspace(6\%)& 5  \enspace(1\%) & 2 \enspace(5\%) & 52 \enspace(2\%) & $< 0.00001$ \\
 jMutRepair   & 7  \enspace(2\%) & 11        (<1\%) & 20 \enspace(5\%)& 24 \enspace(8\%) & 3 \enspace(7\%) & 65 \enspace(3\%) &  $0.009309$ \\
 Nopol        & 1         (<1\%) & 72 \enspace(6\%) & 107      (27\%) & 32        (10\%) & 1 \enspace(2\%) & 213       (10\%) & $< 0.00001$ \\
 DynaMoth     & 0  \enspace(0\%) & 124       (10\%) & 74       (18\%) & 6  \enspace(2\%) & 2 \enspace(5\%) & 206       (10\%) & $< 0.00001$ \\
 NPEFix       & 1         (<1\%) & 3         (<1\%) & 9 \enspace(2\%) & 0  \enspace(0\%) & 2 \enspace(5\%) & 15        (<1\%) & $< 0.00001$ \\
     \midrule
     Total        &    74 &      304 &       550 &            139 &       31 & \numprint{1098} \\
     Total unique & 25 (9.96\%) & 173 (14.93\%) & 187 (47.34\%) & 62 (20.87\%) & 12 (30\%) & 459 (21.44\%) \\
    \bottomrule
    \end{tabular}
\end{table*}

To our understanding, the first hypothesis is more plausible since we observe in RQ1 that the implementation of the repair tools has an impact on their repairability.
However, additional studies should be designed to identify which hypothesis, or a combination of hypotheses, has an impact on the repairability of the repair tools on Defects4J compared to other benchmarks.

\answer{2}{
\textbf{Is the repair tools' repairability similar across benchmarks?}
There is a difference in the repairability of the \nbRepairTools repair tools across benchmarks.
Indeed, the repairability of all tools is significantly higher for bugs from Defects4J compared to the other four benchmarks, therefore we conclude that they overfit Defects4J.
In addition, we raised three hypotheses that might explain this difference.
The confirmation of those hypotheses are full contributions themselves, therefore our study opens the opportunity for several future investigations.
}

\subsection{[RQ3] Causes of Non-patch Generation}

In this final research question, we analyze the repair attempts that did not result in patches, and we identify the causes of non-patch generation.
The goal of this research question is to provide highlights to the automatic repair community on the causes of non-patch generation so that authors of repair tools can improve their tools.

\autoref{tab:repro:exeuctiontimeerror} and \autoref{tab:repro:timeout} present the percentage of repair attempts that finished due to an error and by timeout, respectively.
They show that the repair attempts in error or timeout represent the majority of all repair attempts (56.49\%).
The Bugs.jar benchmark is the main contributor to this percentage.
The size and complexity of the Bugs.jar projects show the limitation of the current automatic patch generation tools.
Moreover, \autoref{tab:repro:exeuctiontimeerror} shows that NPEFix is the tool with the highest error rate, but this tool crashes when no null pointer exception is found in the execution of the failing test case that exposes a bug.
Regarding the timeout in \autoref{tab:repro:timeout}, jGenProg and Cardumen are more subject to reach timeout. 

\begin{table}[t]
    \caption{Percentage of repair attempts that failed by error.}
    \label{tab:repro:exeuctiontimeerror}
    \centering
    \small
    \begin{tabular}{l|rrrrr|r@{}}
    \toprule
\diagbox[height=6em,width=7em]{Repair tool}{Benchmark} & \thead{\rotatebox[x=1.3cm]{90}{Bears}} & \thead{\rotatebox[x=1.3cm]{90}{Bugs.jar}} & \thead{\rotatebox[x=1.32cm]{90}{Defects4J}} & \thead{\rotatebox[x=1cm]{90}{IntroClassJava}} & \thead{\rotatebox[x=1.3cm]{90}{QuixBugs}} & \thead{\rotatebox[x=1.3cm]{90}{Average}} \\
             \midrule
 ARJA        & 24.70 &    49.56 &      1.26 &              0 &        0 &    29.93  \\
 GenProg-A   & 88.04 &    78.06 &      7.08 &              0 &      2.5 &    53.90  \\
 Kali-A      & 24.70 &    50.08 &      4.81 &              0 &        0 &    30.87  \\
 RSRepair-A  & 87.25 &    79.27 &      6.83 &              0 &      2.5 &    54.41  \\
 Cardumen    & 47.41 &    70.46 &     48.60 &              0 &      5.0 &    52.73  \\
 jGenProg    & 45.01 &    63.29 &     12.65 &              0 &      5.0 &    41.94  \\
 jKali       & 44.62 &    64.42 &     12.40 &              0 &      5.0 &    42.45  \\
 jMutRepair  & 72.11 &    66.66 &     15.44 &          13.46 &     22.5 &    49.64  \\
 Nopol       & 28.68 &    60.27 &     45.31 &              0 &      2.5 &    44.37  \\
 DynaMoth    & 27.09 &    46.97 &      4.30 &              0 &        0 &    29.37  \\
 NPEFix      & 89.24 &    86.18 &     73.16 &              0 &      2.5 &    70.62  \\
 \midrule
 Average     & 52.62 &    65.02 &     21.08 &           1.22 &     4.31 &   45.48  \\
    \bottomrule
    \end{tabular}
\end{table}

\begin{table}[t]
\caption{Percentage of repair attempts that failed by timeout.}
    \label{tab:repro:timeout}
    \centering
    \small
    \begin{tabular}{l|rrrrr|c@{}}
    \toprule
\diagbox[height=6em,width=7em]{Repair tool}{Benchmark} & \thead{\rotatebox[x=1.3cm]{90}{Bears}} & \thead{\rotatebox[x=1.3cm]{90}{Bugs.jar}} & \thead{\rotatebox[x=1.32cm]{90}{Defects4J}} & \thead{\rotatebox[x=1cm]{90}{IntroClassJava}} & \thead{\rotatebox[x=1.3cm]{90}{QuixBugs}} & \thead{\rotatebox[x=1.3cm]{90}{Average}} \\
             \midrule
 ARJA        & 19.52 &    18.56 &      6.07 &              0 &        0 &    13.45  \\
 GenProg-A   &  6.37 &     7.08 &      9.62 &              0 &      5.0 &     6.44  \\
 Kali-A      &  1.19 &     2.76 &         0 &              0 &        0 &     1.63  \\
 RSRepair-A  &  7.17 &     6.99 &      8.86 &              0 &      5.0 &     6.35  \\
 Cardumen    &  4.38 &    61.57 &     19.74 &              0 &      2.5 &    37.50  \\
 jGenProg    & 48.20 &    28.23 &     71.39 &          83.83 &     85.0 &    47.31  \\
 jKali       &  0.79 &     4.05 &      1.77 &              0 &        0 &     2.61  \\
 jMutRepair  &  0.39 &     3.45 &      1.01 &              0 &        0 &     2.10  \\
 Nopol       &  0.39 &     0.51 &         0 &              0 &        0 &     0.32  \\
 DynaMoth    &     0 &     0.69 &      2.27 &              0 &        0 &     0.79  \\
 NPEFix      &     0 &     4.49 &      0.75 &              0 &        0 &     2.56  \\
 \midrule
 Average     &  8.04 &    12.58 &     11.04 &           7.62 &     8.86 &    11.01  \\
 \bottomrule
    \end{tabular}
    
\end{table}

We then manually analyzed the execution trace of the repair attempts \cite{Logs} to identify the causes of non-patch generation.
The methodology for this analysis is described in \autoref{sec:repro:patchanalysis}, and by following it we identified six causes of non-patch generation.

\noindent\textbf{1. [The repair tool cannot repair the bug]} A logical problem is that the repair tools do not have a patch that fixes the bug in their search space.
For instance, NPEFix is not able to generate patches for bugs that are not related to null pointer exception.
jGenProg is not able to generate a patch when the repair ingredient is not in the source code of the application, which happens frequently for small programs like the ones in QuixBugs.
New repair approaches should be created to handle this cause of non-patch generation.

\noindent\textbf{2. [Incorrect fault localization]} When the fault localization does not succeed to identify the location of the bug, the repair tools do not succeed to generate a patch for it. 
This can be due to a limitation of the fault localization approach or to the suspiciousness threshold that the repair tools use.
Moreover, we identified that test cases that should pass are failing, and consequently there is a misleading fault localization. 
For instance, the fault localization fails on all bugs from the INRIA/spoon project (from the Bears benchmark) because the fault localization does not succeed to load a test resource, and consequently the passing test cases fail.

\noindent\textbf{3. [Multiple fault locations]} Developers frequently fix a bug at more than one location in the source code: we refer to this type of bug as multi-location bug.
However, most current repair tools and fault localization tools do not support multi-location bugs.
For instance, the bug \href{http://program-repair.org/defects4j-dissection/#!/bug/Math/1}{Math-1} from Defects4J has to be fixed in the exact same way at two different locations, and the two locations are specified by two failing test cases. 
The current tools consider that the two failing test cases specify the same bug at the same location, and consequently do not succeed to generate a multi-location patch.


\noindent\textbf{4. [Too small time budget]} We observe that some of the repair attempts finish the execution by consuming all the time budget.
Considering the size of this experiment, it is not realistic to increase drastically the time budget. 
However, new approaches and optimizations can minimize this problem.
In this study, we detected \numprint{2593} repair attempts that failed by timeout.
It is not possible to predict the outcome of those attempts, but a previous study \cite{Martinez2018Cardumen} showed that additional time budget might result in a higher number of generated patches by genetic programming approaches. 
However, in our experiment, the repair tools require 13.5 minutes on average to generate a patch, which is significantly lower than the allocated time budget (two hours).

\noindent\textbf{5. [Incorrect configuration]} We also observe that the \frameworkName framework does not succeed to correctly compute parameters from some bugs to give as input to the repair tools, such as compliance level, source folders, and failing test cases. This results in failing repair attempts, which can be due to a bug in \frameworkName or an impossibility to compile the bug.
For instance, NPEFix fails 215 times because of issues related to classpath.\footnote{All the occurrences of InvalidClassPathException in our execution:
\url{https://git.io/fjRax}
} 

\noindent\textbf{6. [Other technical issues]} The final cause of non-patch generation is related to other technical limitations that cause the non-execution of the repair tools. 
One of them is about too long command lines. The repair tools are executed from the command line, which means that all parameters must be provided in the command line. However, the size of the command line is limited, and in the case of projects that have a long classpath, the operating system denies the execution of the command line, which results in failing repair attempts.
On Bugs.jar, for instance, 200 repair attempts finished with the error \texttt{[Errno 7] Argument list too long}.\footnote{Repair attempts that end with \texttt{Argument list too long}: 
\url{https://git.io/fjRap}
}
Finally, there are also other diverse issues that cause the repair tools to crash. 
For instance, jGenProg finished its repair attempt on the bug Flink-6bc6dbec from Bugs.jar with a \texttt{NullPointerException}.\footnote{Log file of the repair attempt:
\url{https://git.io/fjRab}
}

\answer{3}{
\textbf{What are the causes that lead repair attempts to not generate patches?}
Through an analysis on logs of repair attempts, we identified six causes of non-patch generation, such as incorrect fault localization.
Each cause should be investigated in detail in new studies.
Moreover, repair tools' designers are also stakeholders on those causes, which inform them what are the weakness of their tools and help them to understand their previous evaluations' results.
}

\section{Discussion}\label{sec:repro:discussion}

\noindent\textbf{Diversity of program repair benchmarks.}
In RQ2, we found that all \nbRepairTools Java repair tools included in this study perform significantly better on the bugs from Defects4J than on the bugs from other benchmarks.
Indeed, repair tool evaluations that only use Defects4J have a threat to the external validity since the repairability results cannot be generalize for other benchmarks.
We then conclude that future tools should be evaluated on diverse benchmarks to mitigate that threat.

\noindent\textbf{Impact of the repair tools' engineering on the repairability.}
During the conduction of this study, we observed that the implementations of the repair approaches play an important role in their ability to repair bugs.
For instance, jKali and Kali-A share the same approach, but they neither have the same implementation nor the same results (see \autoref{tab:repro:nbgeneratedpatches}).
Kali-A fixes 118 bugs while jKali fixes 52 with the same input.
Note that this observation has also been correlated with the analysis of non-patch generation, where a significant number of causes is not related to the repair approaches themselves, but to their implementations.
This observation highlights a potential bias in empirical studies on automatic program repair that compare the repairability of different repair approaches.
Based on this observation, those studies only compare the effectiveness of repair tools, not the approaches themselves.

\noindent\textbf{Challenges of creating \frameworkName.}
The main challenges we faced to run repair tools are related to the creation of the \frameworkName framework.
First, we checked all test-suite-based repair tools for Java based on our criteria (e.g. availability) so that we could find the suitable tools for our study (see \autoref{tab:selection-tools}).
Then, we had to understand the repair tools we finally gathered, where manual source code analysis was required so that we could compile them and find their inputs and requirements: the tools are diverse, sometimes not documented, and implemented by different researchers.
Once we understood the repair tools, we could plug them in the \frameworkName framework, which contains the abstraction around the tools.
Those challenges are mainly due to lack of well-organized open-science repositories for all the repair tools. Good documentation, examples, and instructions on how to compile the tools can speed up the process of learning on how to execute repair tools.

\noindent\textbf{The observed repairability compared to the previous evaluations.}
\autoref{tab:Java-test-suite-based-repair-tools} shows the test-suite-based repair tools for Java and the repairability results from their previous evaluations.
Those results are difficult to compare with the results of our study, because the previous evaluations on Defects4J did not consider all bugs from the benchmark.
On Defects4J, only Cardumen fixes fewer bugs in this study compared to the previous evaluation.
This can be explained by the difference of the setup (such as the number of random seed considered in the study), and potential bugs in the version of Cardumen we use.
According to those results, \frameworkName configures correctly the repair tools to generate patches since no major drawbacks have been observed.



\noindent\textbf{Threats to validity.}
As with any implementation, the \frameworkName framework is not free of bugs. 
A bug in the framework might impact the results we reported in \autoref{sec:repro:results}. 
However, the framework and the raw data are publicly available for other researchers and potential users to check the validity of the results. 

This study focuses on test-suite adequate patches, which means that the generated patches make the test suite pass; yet, there is no guarantee that they fix the bugs.
Studying patch correctness \cite{Xin2017,Yu2019,Le2018overfitting} is out of the scope of this work.
Our goal is to analyze the current state of the automatic program repair tools and identify potential flaws and improvements.
The conclusions of our study do not require the knowledge on the correctness of the patches.

Our goal is to have a full picture of test-suite-based repair tools for Java. In our literature review, presented in \autoref{sec:repro:existing-evaluations-on-repair-tools}, we found \totalRepairToolsLiterature repair tools that compose the full picture. Our study was conducted considering only \nbRepairTools of them: note that this is the largest experiment in terms of number of repair tools (and benchmarks). However, we do not have the full picture we wanted to, which is a threat to the external validity of our results. Most of the repair tools that we did not include in our study are just not possible to be ran: for instance, PAR \cite{Kim2013PAR} is not even available.
Open-source tools allow the community to generate knowledge in several directions. In our work, open-source tools allowed us to perform a novel evaluation on the state of the repair tools. Another direction is to help the development of new tools: for instance, DeepRepair \cite{White2019DeepRepair} is built over Astor \cite{Martinez2016Astor}, which is a library for repairing.

\vspace{-.5em}
\section{Related Works}\label{sec:repro:related-works}


The works related to ours are empirical studies on the \textit{repairability} of multiple automatic program repair tools.
Repair tools for C programs were the subject of investigation in the first empirical studies on automatic program repair.
Qi et al. \cite{Qi2015Kali} introduced the idea of \textit{plausible} (i.e. \textit{test-suite adequate}) versus \textit{correct} patch.
They studied the patch plausibility and correctness of four generate-and-validation repair tools on the bugs from the GenProg benchmark \cite{LeGoues2012study} (which later became a part of the ManyBugs benchmark \cite{LeGoues2015ManyBugsIntroClass}). They found that a small number of bugs are fixed by correct patches.

An \textit{incorrect}, test-suite adequate patch is known as an \textit{overfitting patch}, because it overfits the test suite. Such problem was named as \textit{the overfitting problem} by Smith et al. \cite{Smith2015}, who studied it in the context of two generate-and-validate C repair tools on the bugs from the IntroClass benchmark \cite{LeGoues2015ManyBugsIntroClass}.
They found that even using high-quality, high-coverage test suites results in overfitting patches.
Later, Le et al.~\cite{Le2018overfitting} analyzed the overfitting problem for semantics-based repair techniques for the C language. The study also investigates how test suite size and provenance, number of failing tests, and semantics-specific tool settings can affect overfitting.

For Java, Martinez et al. \cite{Martinez2017experiment} reported on a remarkable large experiment, where three repair tools
were executed on the bugs from Defects4J. The focus of their study was to measure the repairability of the repair tools and to find correct patches by manual analysis.
They also found that a small number of bugs (9/47) could be repaired with a test-suite adequate patch that is also correct.
Ye et al. \cite{Ye2019StudyQuixBugs} presented a study where nine repair tools were executed on the bugs from QuixBugs. They used automatically generated test cases based on the human-written patches to identify incorrect patches generated by the repair tools.

Motwani et al. \cite{Motwani2018evaluation} reported on an empirical study that included seven repair tools for both Java and C languages, where the Defects4J and ManyBugs benchmarks were used.
They had a different focus: they investigated if the bugs repaired by repair tools are hard and important. To do so, they used the repairability data from previous works, and they performed a correlation analysis between the repaired bugs and measures of defect importance, the human patch complexity, and the quality of the test suite.

\autoref{tab:repro:empirical-studies} summarizes the mentioned studies.
The main difference between our study and all the previous ones is the goal. Previous works focused on the \textit{patch overfitting} problem and advanced/correlation analysis between the repairability of tools and bug-related characteristics.
We introduce the \textit{benchmark overfitting} problem, which is investigated in this paper as well as the causes of non-patch generation.
Moreover, the scale of our study is much larger than previous studies, on repair tools and benchmarks.

\begin{table}[t]
    \caption{Empirical studies on repair tools.}
    \label{tab:repro:empirical-studies}
    \centering
    \footnotesize
    \setlength\tabcolsep{4pt} 
    \begin{tabular}{@{}l p{.045\textwidth} r r p{.25\textwidth}@{}}
        \toprule
        Work & Language & \# Tools & \# Bench & Main focus \\
        \midrule
        \cite{Qi2015Kali} & C & 4 & 1 & plausible vs. correct patch \\
        \cite{Smith2015} & C & 2 & 1 & patch overfitting \\
        \cite{Le2018overfitting} & C & 4 & 2 & patch overfitting \\
        \cite{Martinez2017experiment} & Java & 3 & 1 & patch overfitting \\
        \cite{Ye2019StudyQuixBugs} & Java & 9 & 1 & patch overfitting \\
        \cite{Motwani2018evaluation} & Java + C & 7 & 2 & repairability vs. bug-related measures \\
        Ours & Java & \nbRepairTools & \nbBenchmarks & benchmark overfitting+non-patch generation \\
        \bottomrule
    \end{tabular}
\end{table}




\section{Conclusions}\label{sec:repro:conclusions}

In this paper, we presented an empirical study including \nbRepairTools repair tools and \nbBugs bugs from \nbBenchmarks benchmarks. In total, \nbRepairAttempts repair attempts were performed: this is the largest experiment to our knowledge.
The goal of our experiment is to obtain an overview of the current state of the repair tools for Java in practice. For that, we scaled up the previous experiments by considering more benchmarks of bugs, which combined have bugs from 130 projects, collected with different strategies.

We found that the repair tools are able to repair bugs from benchmarks that were not initially used for their evaluations.
However, our results suggest that all repair tools overfit Defects4J.
Finally, we analyzed why the repair tools do not succeed to generate patches: this study resulted in six different causes that can help future development of repair tools.

Our study opens several opportunities for future investigations. First, our hypotheses on why the repair tools perform better on Defects4J can be further confirmed. For instance, one of the hypotheses is the fact that the buggy program versions were changed in Defects4J due to the bug fix isolation. A study to confirm this hypothesis is a full contribution itself.
Second, other repair tools can also be executed to aggregate and scale up our study. ssFix, for instance, is possible to run, despite the fact we had issues for it, which lead to its exclusion in this work. Moreover, the tools that are hardcoded to be ran on Defects4J could also be adapted to work for other benchmarks of bugs.
Finally, an investigation on the bug type distribution in the benchmarks should also be conducted. This would provide the information on how many bugs a repair tool is actually able to fix, i.e. by finding the bugs that meet the repair tools' bug class target.

\begin{acks}
We acknowledge CAPES for partially funding this research, and Marcelo Maia for discussions.
This material is based upon work supported by Funda\c{c}\~ao para a Ci\^encia e a Tecnologia (FCT), with the reference PTDC/CCI-COM/29300/2017 and  UID/CEC/50021/2019.
\end{acks}

\clearpage

\bibliographystyle{ACM-Reference-Format}
\bibliography{references}

\end{document}